\documentclass[fleqn,usenatbib]{mnras}

\usepackage{newtxtext,newtxmath}

\usepackage[T1]{fontenc}
\usepackage[utf8]{inputenc}
\usepackage{ae,aecompl}

\usepackage{graphicx}   
\usepackage{amsmath}    
\usepackage{amssymb}    
\usepackage{multirow}
\usepackage[usenames,dvipsnames]{color}

\title[Colour spread in M\,3 1G]{Is helium the key parameter in the extended color spread of the first generation stars in M3?}

\author[Tailo, M. et al.]{
M. Tailo$^{1}$,\thanks{E-mail: marco.tailo@unipd.it; mrctailo@gmail.com}
F. D'Antona$^{2}$,
V. Caloi,$^{3}$,
A. P.  Milone$^{1}$,
A. F. Marino$^{1,4}$,
E. Lagioia$^{1}$,
\newauthor
G. Cordoni$^{1}$
\\
$^{1}$Dipartimento di Fisica e Astronomia ``Galileo Galilei'', Univ. di Padova, Vicolo dell'Osservatorio 3, Padova, IT-35122\\
$^{2}$INAF -- Osservatorio Astronomico di Roma, via di Frascati 33, Monte Porzio Catone,  IT-00078 \\
$^{3}$INAF -- IASF Roma, Via Fosso del Cavaliere, Roma, Italy, IT-00133  \\
$^{4}$Centro di Ateneo di Studi e Attivita’ Spaziali “Giuseppe Colombo” - CISAS, Via Venezia 15, Padova, IT-35131}
\date{ Accepted 2019 May 3. Received 2019 May 3; in original form 2019 March 15}

\pubyear{2018-2019}

\begin{document}
\label{firstpage}
\pagerange{\pageref{firstpage}--\pageref{lastpage}}
\maketitle

\begin{abstract}
The study of the "chromosome maps" of Galactic Globular Clusters has shown that the stars identified as `first generation' often define an extended sequence in the $m_{F275W}-m_{F814W}$ colour, whose straightforward interpretation, by comparison with synthetic spectra, is that they are inhomogeneous in helium content. 
The cluster M\,3 (NGC\,5272) is one of the most prominent example of this phenomenon, since its first generation is distributed on an extended colour range, formally corresponding to a large helium enhancement ($ \sim 0.1$).  It is necessary to ask whether the bulk of photometric observations available for this cluster supports or falsifies this interpretation. For this purpose, we examine the horizontal branch morphology, the  period and magnitude distributions of the RR Lyrae variables, and the main sequence colour distribution. Simulating the first generation stars with such internal variation of helium content we can not meet all the observational constraints at the same time, concluding that the origin of the first generation colour spread is still without a straightforward explanation.
\end{abstract}

\section{Introduction}
\label{SEC_itro}

\begin{figure*}
\centering
\includegraphics[width=2\columnwidth]{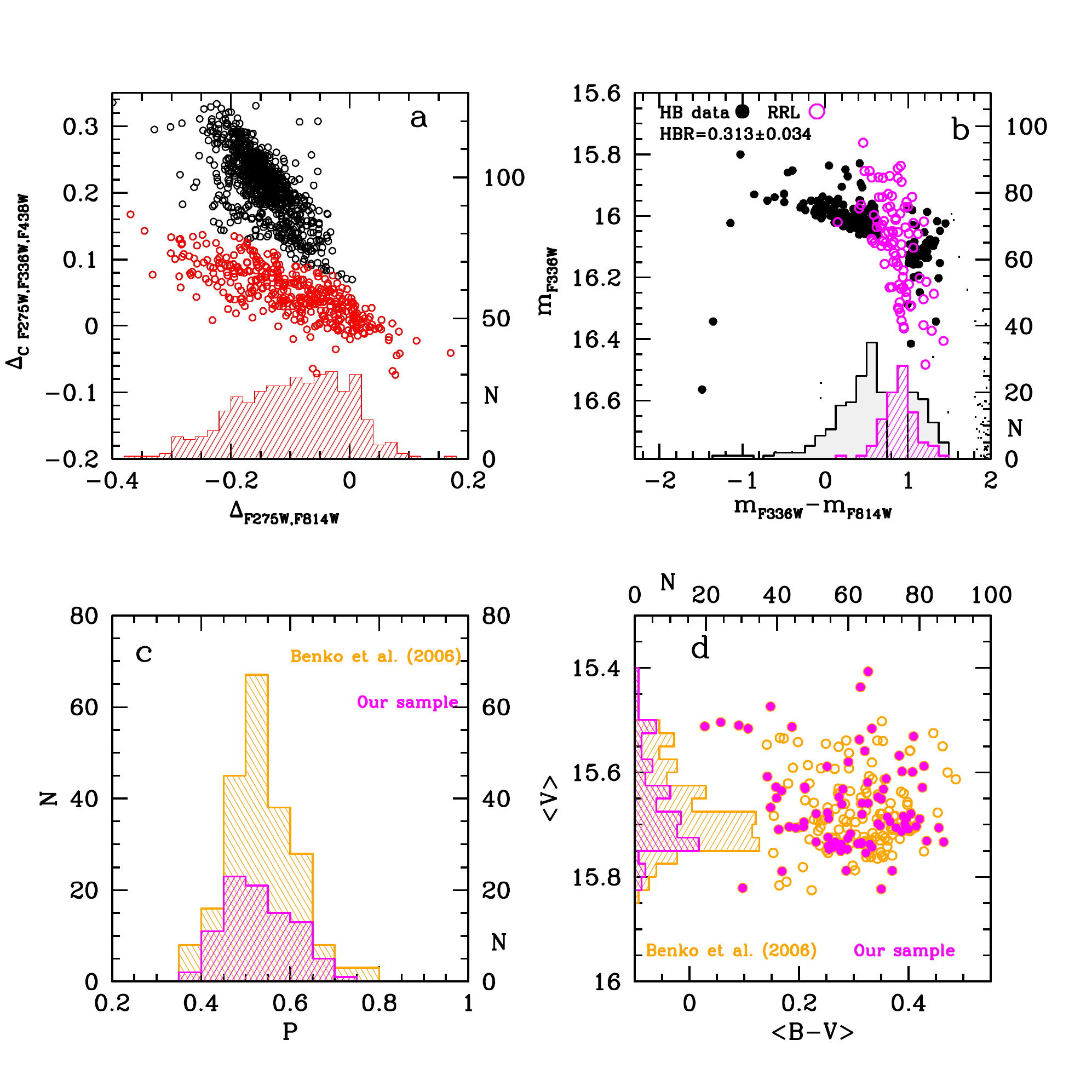}
\caption{
\textit{Panel a:} The ChM of M3 with highlighted the 1G (red) and its colour distribution. 
\textit{Panel b:} The HB subsample in the \textit{HST} data. We identify the RRL variable stars and represent them as magenta, open circles while other HB stars are plot as black, filled ones. The two histograms describe the distribution of the HB stars (black) and of the variable stars (magenta). See text for more details (\S~\ref{SEC_data})
\textit{Panel c:} The period distribution of the RRL variables from the global catalogue \citep[][orange histogram]{benko_2006} and of the ones in the \textit{HST} sample (magenta histogram). The peak at $\rm P\sim0.53$ is still present, but is not as sharp as the one in the general distribution.
\textit{Panel d:}  The colour magnitude diagram,  built from the mean magnitudes, of the RRL variables in the global catalogue \citep[][orange]{benko_2006}. We highlighted the RRL  of the \textit{HST} sample (magenta).  }
\label{PIC_results1}
\end{figure*}

The rich mine of the \textit{Hubble Space Telescope (HST)} UV Legacy Survey of Globular clusters data \citep[GC,][]{piotto_2015} allowed to build a ``chromosome map" (ChM) for each cluster, by plotting for each star specific combination of UV and  optical--near infrared HST bands \citep[a detailed description of the procedure used can be found in][]{milone_2017}. These maps constitute a powerful spectro-photometric tool to distinguish different stellar populations in GCs, but they have unveiled patterns that lack a satisfying explanation in the context of GCs evolution. A prominent example is the fact that \cite{milone_2015,milone_2017} first, then \cite{lardo_2018} and \cite{milone_2018} found that the first generation of star (1G), as defined by the ChMs, in several cases does not look compatible with a homogeneous population.

In fact, the 1G stars show a dispersion in the $m_{F275W}-m_{F814W}$ colour that can be associated to variations of both chemical and physical properties, due to the high sensibility of the special combination of filters employed. This puzzling result, first found during the analysis of the GC NGC2808 \citep{milone_2015,dantona_2016}, and then confirmed for many other clusters by \cite{milone_2017,milone_2018}, leaves us with the doubt that the picture of what could have been the evolutionary process that led to the formation of a GC was even more complex and confusing than we might have envisioned until now.
  
While the second generation (2G) in the ChMs can be interpreted as the result of CNO burning \citep{milone_2017}, the 1G stars distribution should have a different origin.
\cite{milone_2015} and \cite{lardo_2018} tentatively attributed it to pure helium differences among the 1G stars. Although following similar comparison with synthetic spectra, finding maximum values of helium enhancements ranging from  $\sim0.00$ to $\sim0.10$, \cite{milone_2018} were very doubtful on the feasibility of this explanation, based on simple theoretical grounds.  
Nevertheless, recently \cite{cabrera_2019} have obtained the abundances of C, N, O, Na, Mg and Al of stars all along the extended 1G of NGC\,2808, and found that they are all homogeneous in these elements, reinforcing the points made by \cite{milone_2015} using multiband photometry; this recent work then gives strenght to the proposal that the color extension of the 1G is due to pure helium enhancement.  Furthermore, \cite{marino_2019}, performing a combined examination of the ChM and the spectroscopic data of a large number of GCs, found the same kind of homogeneity in multiple cluster. Their interpretation, on the other hand, is twofold and in addition to helium enhancement they also attribute the extended 1G sequence to iron variations.
Therefore, in the absence of direct helium determination,  it is important to find signatures of such a postulated helium variation by examining its influence on the location and properties of other stars in the color magnitude diagram

The cluster NGC\,5272 (M 3) is a perfect example of an extended 1G \citep[see its ChM in ][]{milone_2017,milone_2018}. We examine its color magnitude diagram in order to investigate whether the interpretation of the 1G extension as a result of a pure helium variation is consistent with the available observations for this GC and, most importantly, its horizontal branch (HB) and variable stars.

The stars' distribution along the HB has indeed been examined at length in the literature  \citep[e.g.][]{catelan_2004,castellani_2005,caloi_2008,denissenkov_2017}.
Almost unique among the galactic GCs,  M\,3 has a large  (>200) population of RR Lyrae variables \citep[RRL,][]{corwin_2001,benko_2006}. The high sensitivity of the period distribution of these variables to any variations of the stellar parameters makes them a great probing tool for this kind of investigation. 

We will achieve our goal comparing state of the art stellar population models with the most recent observations available. In doing so we will also produce a new and updated description of the HB in M3.

 The present work can be divided into two main parts. In \S~\ref{SEC_datamod} we present a summary of the observations we exploit and a description of the stellar evolution models we employ. In \S~\ref{SEC_simres}, \S~\ref{SEC_sim2} and  \ref{SEC_SimMS} we present and discuss the results of our simulations. \S~\ref{SEC_Concl} will host our conclusions and general remarks. 

\section{Dataset and Models}
\label{SEC_datamod}
We exploit photometry from the recent HST UV photometric survey \citep{piotto_2015} and from the ACS survey of Galactic GCs \citep{anderson_2008}. Moreover, we used the ChM derived by \cite{milone_2017}. We also employ the RRL variables database by \cite{benko_2006}. We refer to these works for the details of the data acquisition and reduction process, and the procedure used to derive the ChM of this cluster.

\subsection{Summary of the observations}
\label{SEC_data}

Analysing the photometric data, the ChM and its related features, \cite{milone_2018} have been able to estimate the average  and the maximum value of helium enhancement between the 1G and the 2G stars ($\rm  \Delta Y_{2G,1G}$). They found $\rm \Delta Y_{2G,1G}=0.016\pm 0.005$ and $\rm  \Delta Y_{2G,1G}^{max}=0.041\pm 0.009$ \citep[see Table 4 in][]{milone_2018}. These values will be our first observational constraints when updating the description of the HB of this cluster. 

Fig. \ref{PIC_results1}a plots the ChM of M\,3 from \citet[][Figure 5]{milone_2017}, where the 1G and the 2G stars are shown as red and black dots, respectively. 
Figure 8 in \cite{milone_2018} shows that each element plays a distinct role in affecting the position on the ChMs. The 2G can be described as a CNO sequence, as confirmed by the spectroscopy of stars at different locations on the ChM. The prominent extension in $\rm \Delta_{F275W,F814W}$\, of the 1G of M\,3 can be described by a helium enhancement sequence \citep{milone_2015, lardo_2018, milone_2018}, and would correspond to a maximum star-to-star difference in the 1G helium mass fraction $\Delta \rm Y_{1G}^{max} \sim 0.10$ \citep{milone_2018}.
 
Figure \ref{PIC_results1}b shows the  $\rm m_{F336W}$\ versus  $\rm m_{F336W}-m_{F814W}$\ colour magnitude diagram of the HB sample from the \textit{HST} photometric catalogue. The magenta dots highlight the RRL variables. M3 has an extended HB,  stars populate both the red and the blue side of the instability strip. 
In the \textit{HST} sample we count a total of $290\pm 17$.0 HB stars divided in $54\pm 7.4$ red HB stars, $144 \pm 12.0$ blue HB stars and $92 \pm 9.6$ RRL variables\footnote{To each of these number we associate the error estimated from the Poisson distribution. }. The black shaded histogram shows the colour distribution of the non variable stars. 
The HB ratio\footnote{This ratio is defined as HBR=(B-R)/(B+V+R) where B,R and V are the number of blue, red and variable stars respectively.} in the \textit{HST} sample is $\rm HBR\sim 0.313 \pm 0.034$, significantly larger than the ratio obtained from ground based observations covering also the external parts of the cluster \citep[$\rm HBR \sim 0.08$,][]{buonanno_1994,ferraro_1997,catelan_2001}. This difference corroborates the evidence that blue-HB stars are more-centrally concentrated than the red HB \citep{catelan_2001,lee_2019}. This feature is expected if the blue HB is populated by 2G stars, born more concentrated in the cluster core, as envisioned by cooling--flow formation models \citep[e.g.][]{dercole_2008} for clusters having a long relaxations time, where the stars are not fully spatially mixed \citep{vesperini_2013}. 
In this work we focus our investigation on the colour distribution of the non variable HB stars and the value of HBR of the \textit{HST} sample because, in these data, the RRL variables are observed at random phases. If we want to use them to further constrain the properties of the HB stellar populations we then need to use other databases.

As already shown in \cite{corwin_2001} and \cite{benko_2006} both the period distribution and the average magnitude distribution of the RRLs have distinctive features. The period distribution of the $215$ RRL in the \cite{benko_2006} sample is plotted in Fig \ref{PIC_results1}c  (orange histogram), showing a prominent peak at $\rm P\sim 0.53$\,d. 
We identified the variables in common between our \textit{HST} sample and the \cite{benko_2006} sample, and plot their period distribution (magenta histogram). The peak at $\rm P\sim 0.53$\,d  is still present, although less sharp than in the whole sample. Doing a Kolmogorov - Smirnov (KS) test we obtain a value of $\rm p \sim 0.88$, signalling that the two distributions are indeed compatible. 

We show in Fig. \ref{PIC_results1}d the <V> magnitude versus <B--V> colour distribution of the whole RRL sample (orange) and of the  \textit{HST} sample. In both cases, the thickness of the instability strip is $\sim 0.20$ magnitudes \citep{corwin_2001,benko_2006}, and the dimmer part of the magnitude range (V>15.6) is most ($\sim 80\%$) populated. This feature was used to obtain insights on the helium content \citep[e.g as in][]{caloi_2008,denissenkov_2017} with the results that most RRLs have been assigned to the 1G. In this case the KS test gives us a value of $\rm p \sim 0.56$ signalling that also the two magnitude distributions are compatible

The period  and mean magnitude distributions of the RRL variable stars are sensitive to any variations of the stellar parameters ---in particular to possible enhancements in helium mass fraction; thus both will be an invaluable tool when we will consider the possibility that the 1G holds a large internal helium spread.

\begin{table}
\centering
\begin{tabular}{cccccc}
\hline
\hline
&Y&$\rm \sigma^Y$&N&$\rm \mu/M_\odot$&$\rm \sigma^\mu/M_\odot$\\
\hline
Sim.1 &\multicolumn{5}{c}{\small{1G without helium spread}}\\
\hline
1G& 0.250& 0.000& 130& 0.188& 0.005\\
2G$\rm _A$& 0.264& 0.006& 133& 0.204& 0.005\\
2G$\rm _B$& 0.280& 0.006&     7& 0.220& 0.005\\
2G$\rm _C$& 0.291& 0.006&     4& 0.240& 0.005\\
\hline
\multirow{2}{*}{Sim.2} &\multicolumn{5}{c}{\small{1G with large helium spread (Fig. \ref{PIC_results4})}}\\
&\multicolumn{5}{c}{2G: as in Sim. 1}\\
\hline
1G&\multicolumn{2}{c}{0.25$\div$0.35} &130 &0.188  &0.005\\
\hline
\multirow{2}{*}{Sim.2.2} &\multicolumn{5}{c}{\small{1G with large helium spread \citep[as in][]{lardo_2018} }}\\
&\multicolumn{5}{c}{2G: as in Sim. 1}\\
\hline
1G&\multicolumn{2}{c}{0.25$\div$0.28} &130 &0.188  &0.005\\
\hline
\multirow{2}{*}{Sim.3}&\multicolumn{5}{c}{\small{1G with large helium spread  (Fig. \ref{PIC_results4}) and lower mass loss}}\\
&\multicolumn{5}{c}{2G: as in Sim. 1}\\
\hline
1G &\multicolumn{2}{c}{0.25$\div$0.35} & 130 &0.175 &0.005\\
\hline
\multirow{3}{*}{Sim.4}& \multicolumn{5}{c}{\small{Independent RRL}}\\
&\multicolumn{5}{c}{\small{Non variable 1G stars with large helium spread (Fig. \ref{PIC_results4})}}\\
&\multicolumn{5}{c}{2G: as in Sim. 1}\\
\hline
1G NV        &\multicolumn{2}{c}{0.25$\div$0.35}&55	  &0.189 &0.005\\
1G RRL &0.250  &0.000 &75	  &0.189 &0.005\\
\multicolumn{6}{c}{2G: as in Sim. 1}\\
\hline
\multirow{4}{*}{Sim.5}&\multicolumn{5}{c}{\small{Independent RRL}}\\
&\multicolumn{5}{c}{\small{Non variable 1G stars with}}\\
&\multicolumn{5}{c}{\small{ large helium spread (Fig. \ref{PIC_results4}) and lower mass loss }}\\
&\multicolumn{5}{c}{2G: as in Sim. 1}\\
\hline
1G NV &\multicolumn{2}{c}{0.25$\div$0.35}    &55	    &0.171  		&0.005\\
1G RRL &0.250  &0.000 &75	    &0.189  		&0.005\\
\hline
\hline
\end{tabular}
\caption{
The input parameters used to obtain the simulations shown in this work. Columns are:  the value of helium mass fraction and its spread (Y and $\rm \sigma^Y$), the number of stars in each group and the value of mass loss with its spread ($\mu$ and $\sigma^\mu$). A tag on the leftmost column identifies if the group of stars belongs to the 1G or the 2G and if it used for the RRL or for the red non variable stars (NV). For Sim. 2 to 5, the helium mass fraction distribution of the 1G stars is described in \S~\ref{SEC_sim2}
}
\label{TAB_inputs}
\end{table}

 \begin{table*}
 \centering
 \begin{tabular}{ccccccccccc}
 \hline
 \hline
 &HBR&$\rm \bar{M}_{HB}^{1G}/M_\odot$&$\rm \bar{M}_{HB}^{2G}/M_\odot$&$\rm \delta \bar{Y}_{1G}$ & $\rm \delta \bar{Y}_{2G}$ &$\rm \bar{P}_{RRL}/days$&$\rm \bar{M}^{RRL}_V$& $\rm p_{KS1}$& $\rm p_{KS2}$& $\rm p_{KS3}$\\
 \hline
 Sim.1&$0.322$&$0.659$&$0.621$&$\sim 0.000$&$0.016$&$0.532$&$15.66$&0.87&0.84&0.92\\
 Sim.2&$0.649$&$0.624$&$0.621$&$0.026$&$0.016$&$0.507$&$15.63$&$<<0.001$&0.09&0.03\\
 Sim.2.2&$0.462$&$0.651$&$0.621$&$0.018$&$0.016$&$0.505$&$15.64$&$0.05$&0.02&$<0.001$\\
 Sim.3&$0.517$&$0.636$&$0.621$&$0.026$&$0.016$&$0.521$&$15.61$&0.02&0.04&$<0.001$\\
 Sim.4&$0.506$&$0.641$&$0.621$&$0.013$&$0.016$&$0.525$&$15.67$&0.007&0.56&0.77\\
 Sim.5&$0.384$&$0.651$&$0.621$&$0.013$&$0.016$&$0.532$&$15.66$&0.60&0.58&0.95\\
 \hline
 \hline
 \end{tabular}
\caption{
 Some general parameters of the simulations shown in \S~\ref{SEC_simres} and \ref{SEC_sim2}. Columns are: the value of HBR, the mean value of mass ($\rm M_{HB}$) and helium enrichment ($\rm \delta Y$) for both the 1G and the 2G stars,  the mean value of the RRL period ($\rm P_{RRL}$), the mean V magnitude of the RRL variable and the p values from the three KS test we perform during our analysis.
 }
 \label{TAB_results}
 \end{table*}

\begin{figure}
\centering
\hspace{-0.2cm}
\includegraphics[width=1.05\columnwidth]{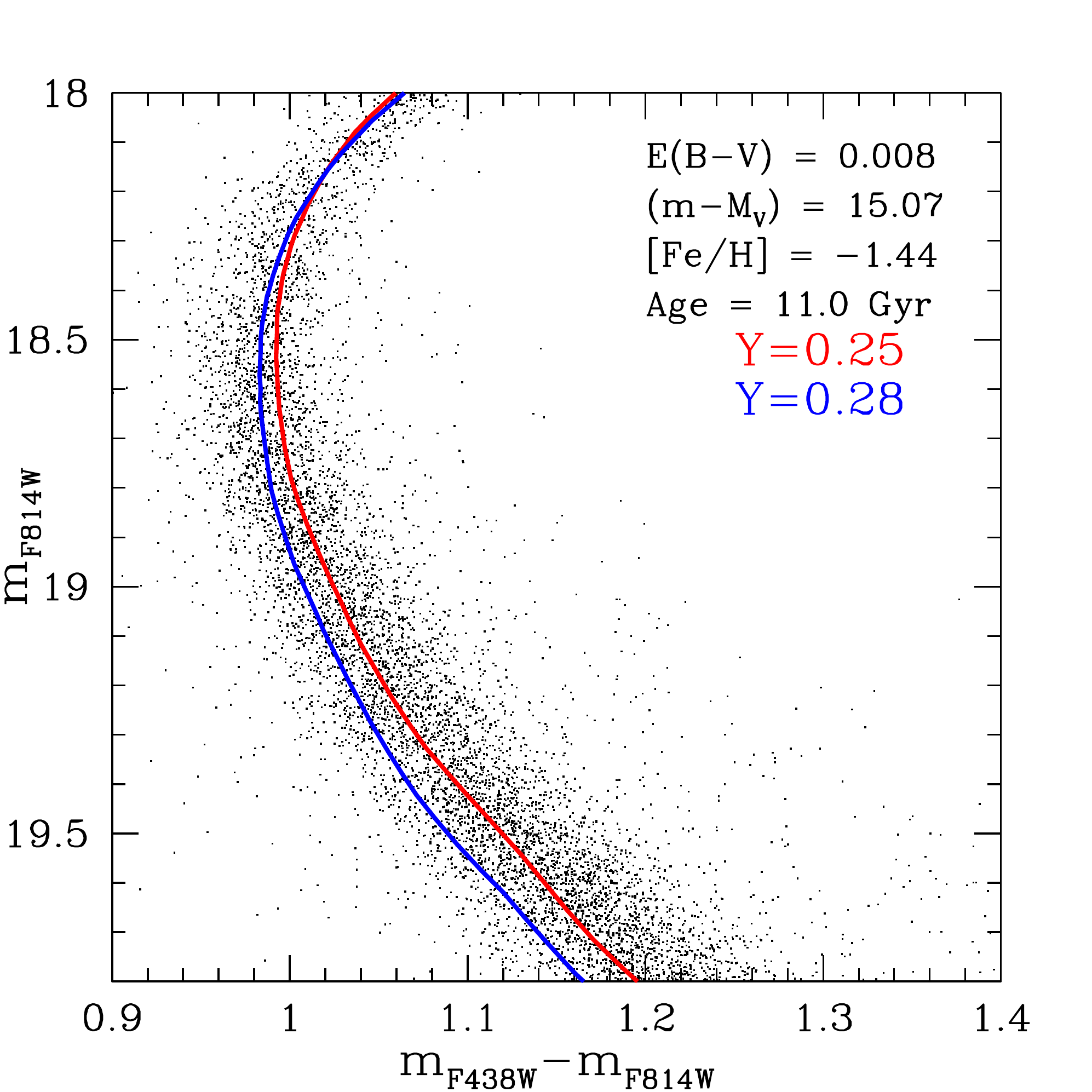}
\caption{
Isochrone fitting in the $\rm m_{F814W}$ vs $\rm (m_{F438W}-m_{F814W})$ CMD of MS stars in M3. We reported our best-fit value of E(B-V), $\rm M_V$ and age.
}
\label{PIC_results2}
\end{figure}

\subsection{Models and simulations }
\label{SEC_models}

The models for this work were developed in \cite{tailo_2015,tailo_2016b,tailo_2017}. While we refer to \cite{tailo_2016b} for the detailed description, we remind the reader the inputs necessary to better understand the present work.
The stellar evolutionary tracks have been calculated with the stellar evolution code ATON2.0 \citep{ventura_1998,mazzitelli_1999}, including the most recent updates for the physical inputs. The HB tracks and population synthesis models have been obtained following the recipes of  \cite{dantona_2002,dantona_2005} and \cite{caloi_2008}.

Briefly, we fix the mass of the each star ($\rm M^{HB}$) as follows: $\rm M^{HB}=M_{Tip}(Z, Y, A) - \Delta M(\mu,\sigma^\mu)$. Here $\rm M_{Tip}$ is the mass at the red giant branch (RGB) tip, function of age (A), metallicity (Z) and helium (Y); $\rm \Delta M$ is a Gaussian describing the mass lost by the star during the RGB phase, $\mu$ and $\sigma^\mu$ its central value and the standard deviation. The values of $\rm M_{Tip}$ are obtained from the isochrone database from \cite{tailo_2016b}. 
When we simulate a HB stellar population with helium mass fraction spread we assume that the helium content is described by a Gaussian distribution, where Y and $\rm \sigma^Y$ are the central value and the standard deviation, respectively.
When calculating the properties of those stars that cross the instability strip and become RRL variables we use the prescription of \cite{marconi_2015}. In simulating the mean <V> and <B> magnitude and colours of the RRL variables, we introduce the errors estimated in \cite{benko_2006}.

We select the set of models having Z=0.001 and $\rm [ \alpha /Fe]=0.4$, corresponding to $\rm [Fe/H] \sim -1.44$ for $Z_\odot =0.014$ \citep[see][]{asplund_2009}. The choice is in agreement with [Fe/H] = --1.50 given by \cite{sneden_2004} and \citet[][as updated in 2010]{harris_1996} for M\,3.

We first fit the MS in order to fix the reddening, E(B-V), distance modulus, $\rm (m-M_{V})$, and age. With our choice of isochrones we obtain: E(B-V)=0.008, $\rm (m-M_{V})=15.07$ and $\rm Age=11.0$\,Gyr (Figure \ref{PIC_results2}).

We compare the data of the HB stars and RRL variables with a large array of simulations, arranged on a grid, where the parameters we can not constrain directly from the observations are progressively changed. In detail, the grids are built by varying the mass loss in steps of 0.001 $\rm M_\odot$: we changed the mass loss of the 1G stars from  $\mu_{\rm 1G}=0.160$ to $\rm 0.220\ M_{\odot}$ and the mass loss of the 2G samples from $\mu_{\rm 2G}=0.180$ to $\rm 0.260\ M_{\odot}$. The mass loss spread for the 1G ($\rm \sigma^\mu_{1G}$) and/or for the 2G samples ($\rm \sigma^\mu_{2G}$) is changed in steps of 0.001 $\rm M_\odot$ from 0.000 to 0.010 $\rm M_\odot$. This allows to constrain quantitatively the difference in mass loss between the different groups. We assess the quality of each simulation in the grid both comparing the histograms in the figures and performing a series of KS tests.

Once we have analysed the HB data of this cluster, we will test if our findings are adequate to describe the MS as well. To do so, we perform a series of MS simulations following the recipes described in \cite{tailo_2016b}. The procedure requires only the assumption of an exponent for the mass function ($\zeta$). In this case we adopt $\zeta=-0.7$, the same exponent \cite{tailo_2016b} used for $\omega$ Centauri. Photometric errors have been evaluated with the artificial star procedure developed in \cite{anderson_2008}. We also introduce in the simulation the binary fraction found for M3 by \citet[][$\sim 3.5\%$]{milone_2012}. 

\begin{figure*}
\centering
\includegraphics[width=2\columnwidth]{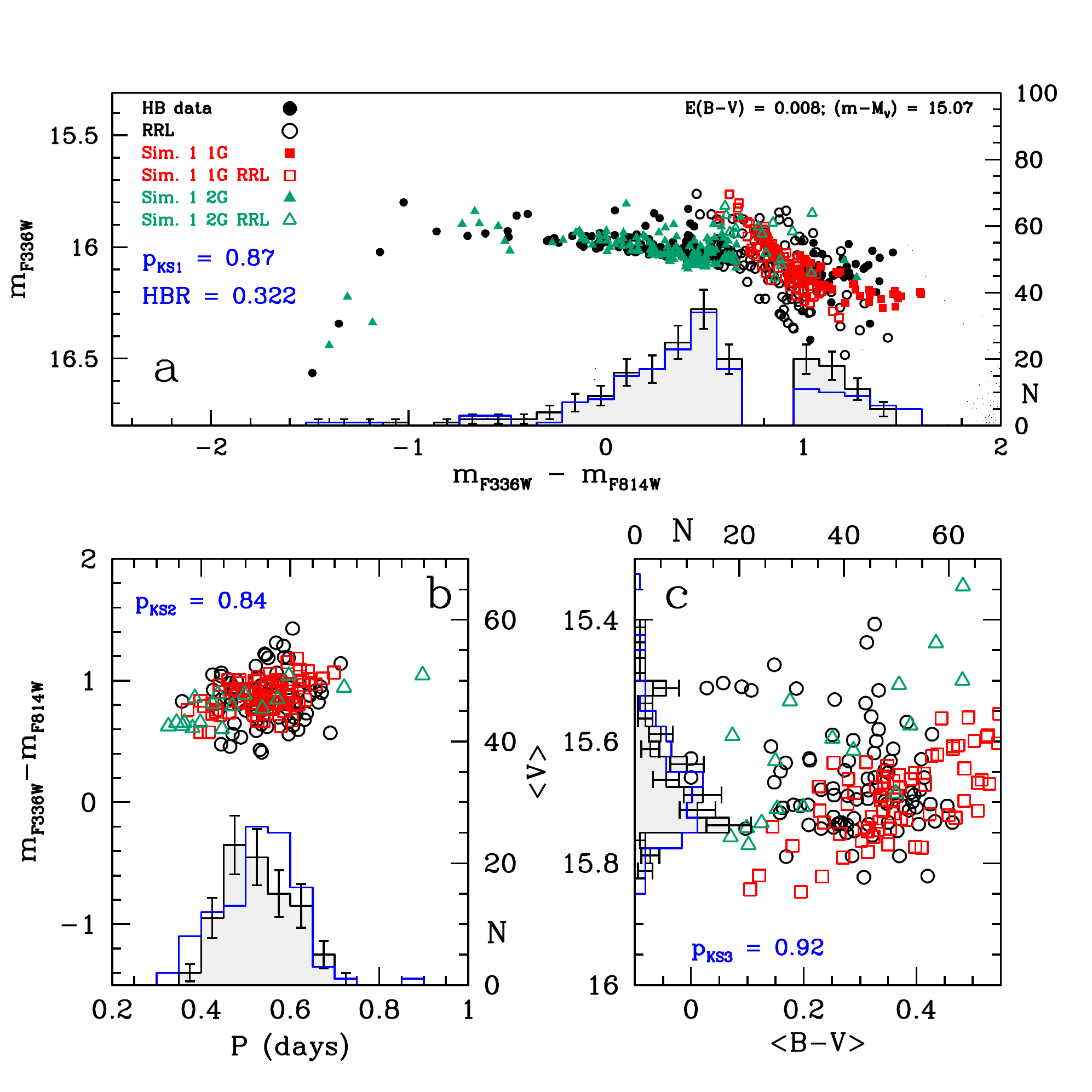}
\caption{
\textit{Panel a:} The results of Sim.1. In the panel, the data are represented as black circles (solid for the HB stars and open for the variables) while the 1G and the 2G stars present in the simulation are the red square and the green triangles, respectively. We represent the variables of the simulation with open points as well. The two histograms, shaded black and blue, respectively for the data and the simulation, describe the agreement between the two sets of points. We reach a satisfying reproduction of the HB morphology. The final inputs used are listed in Table \ref{TAB_inputs}.
\textit{Panel b:}The RRL variables in the \textit{HST} sample and in Sim.1 observed in the P vs  ($\rm  m_{F336W} - m_{F438W}$) plane. The histograms in the panel describe their period distribution. 
\textit{Panel c:} The CMD obtained from the average B and V magnitude of the simulated and the observed RRL variables. The histograms in the panel describe the V magnitude distribution of both samples.
 }
\label{PIC_results3}
\end{figure*}

\begin{figure}
\centering
\includegraphics[width=\columnwidth]{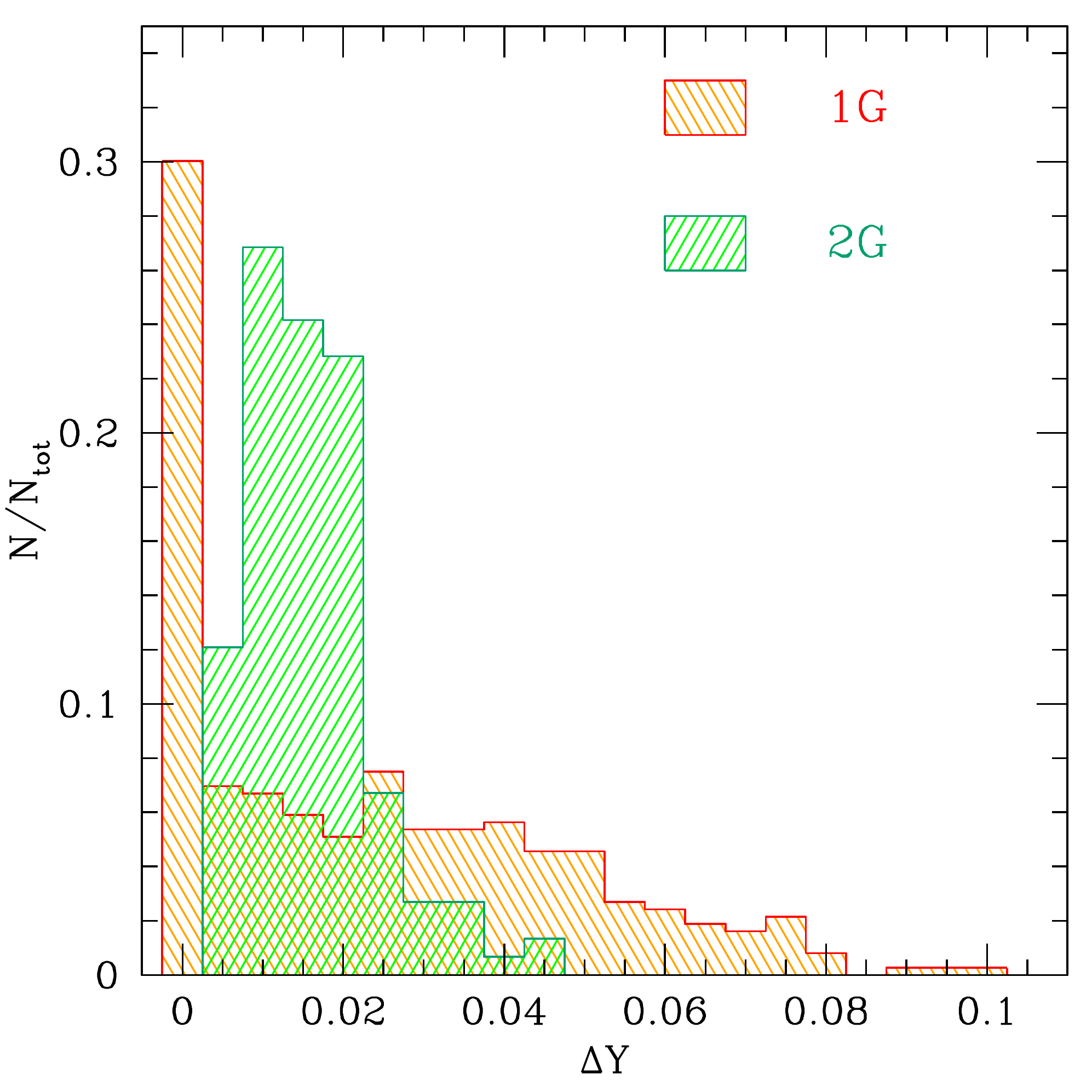}
\caption{
The distribution of helium enhancement ($\rm \Delta Y$) of the 1G (red) and 2G stars (green) calculated in this work. See text for further details.} 
\label{PIC_results4}
\end{figure}

\begin{figure*}
\centering
\includegraphics[width=2\columnwidth]{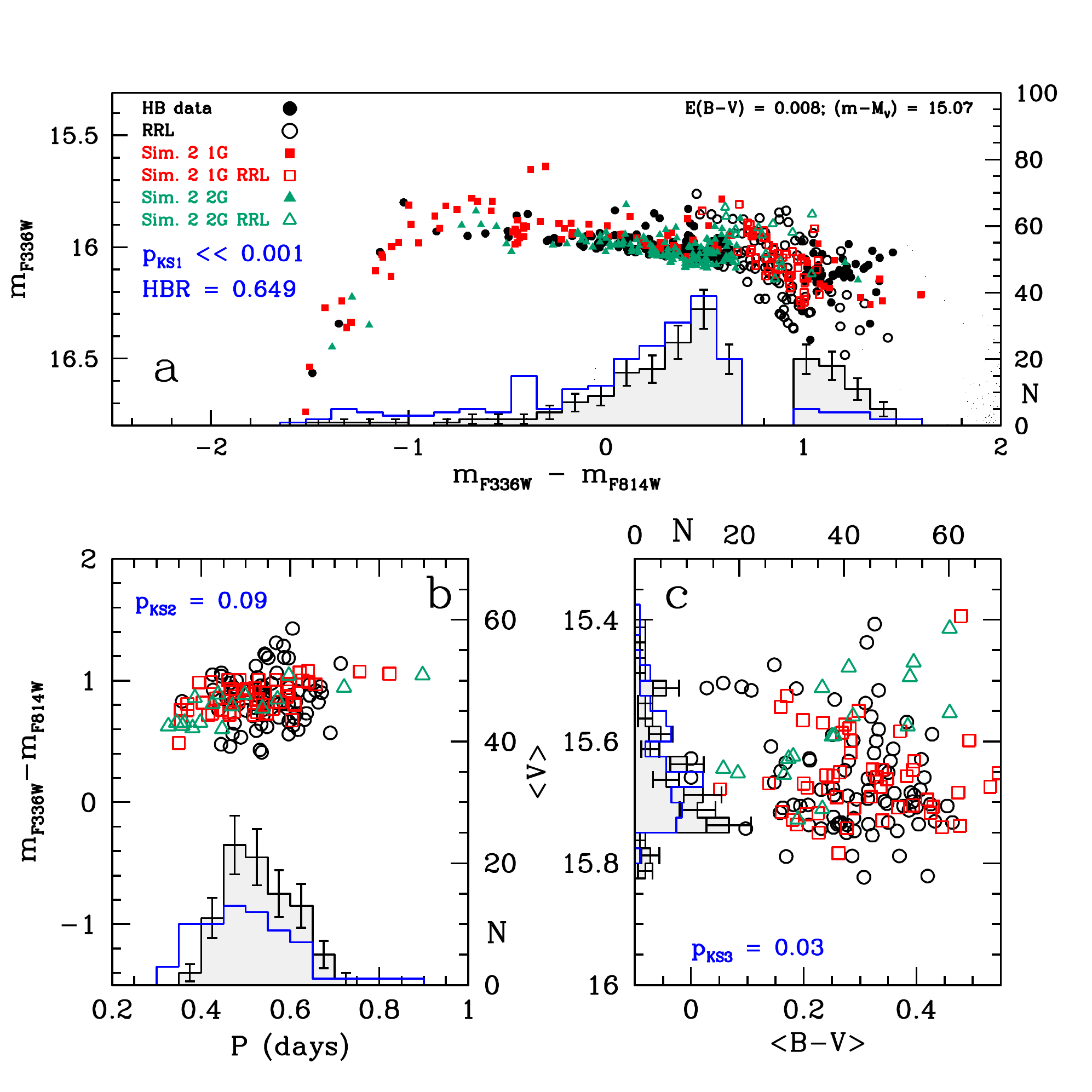}
\caption{
As Figure \ref{PIC_results3} but for Sim. 2. 
}
\label{PIC_results5}
\end{figure*}

\begin{figure*}
\centering
\includegraphics[width=2\columnwidth]{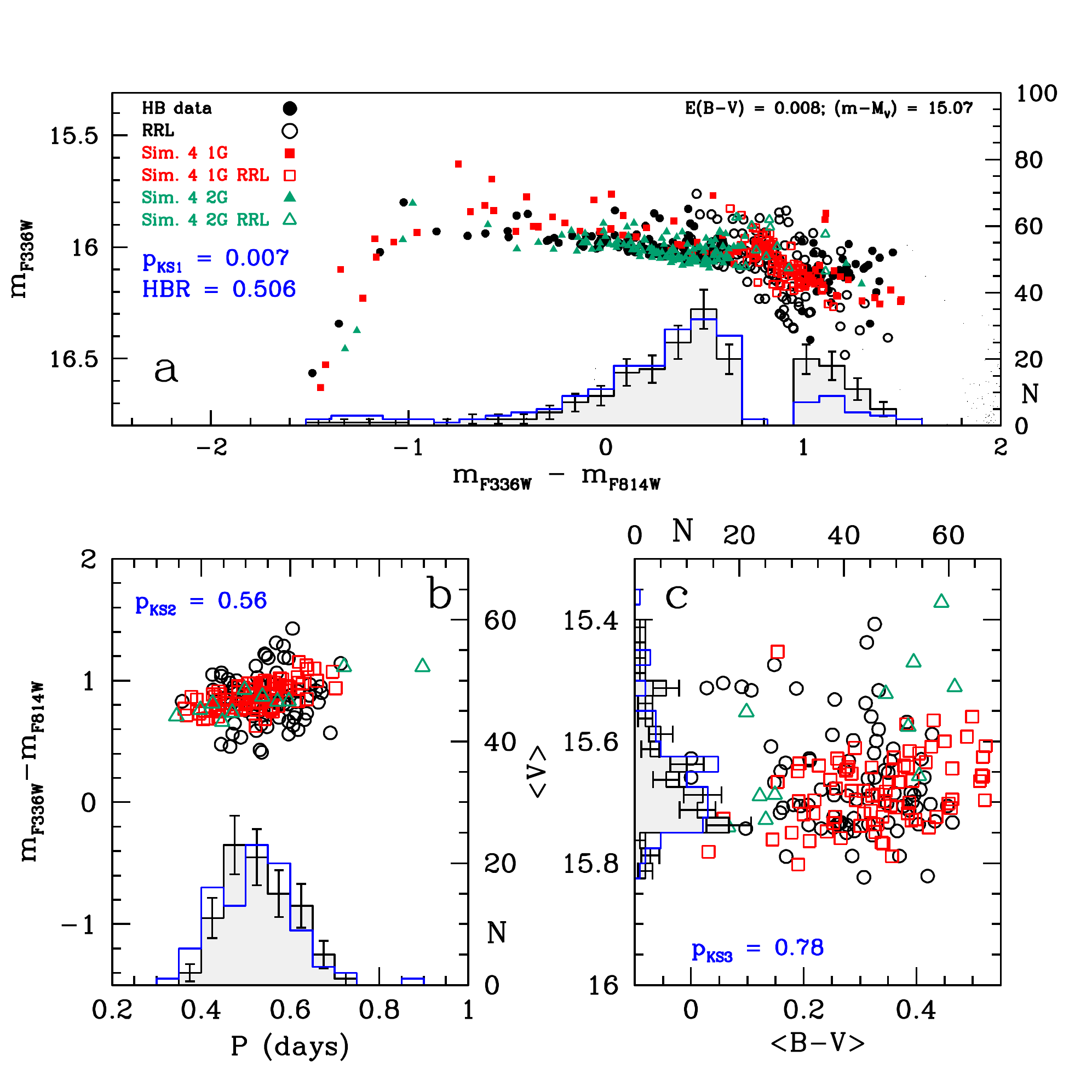}
\caption{
As Figure \ref{PIC_results3} but for Sim. 4. 
}
\label{PIC_results6}
\end{figure*}

\begin{figure*}
\centering
\includegraphics[width=2\columnwidth]{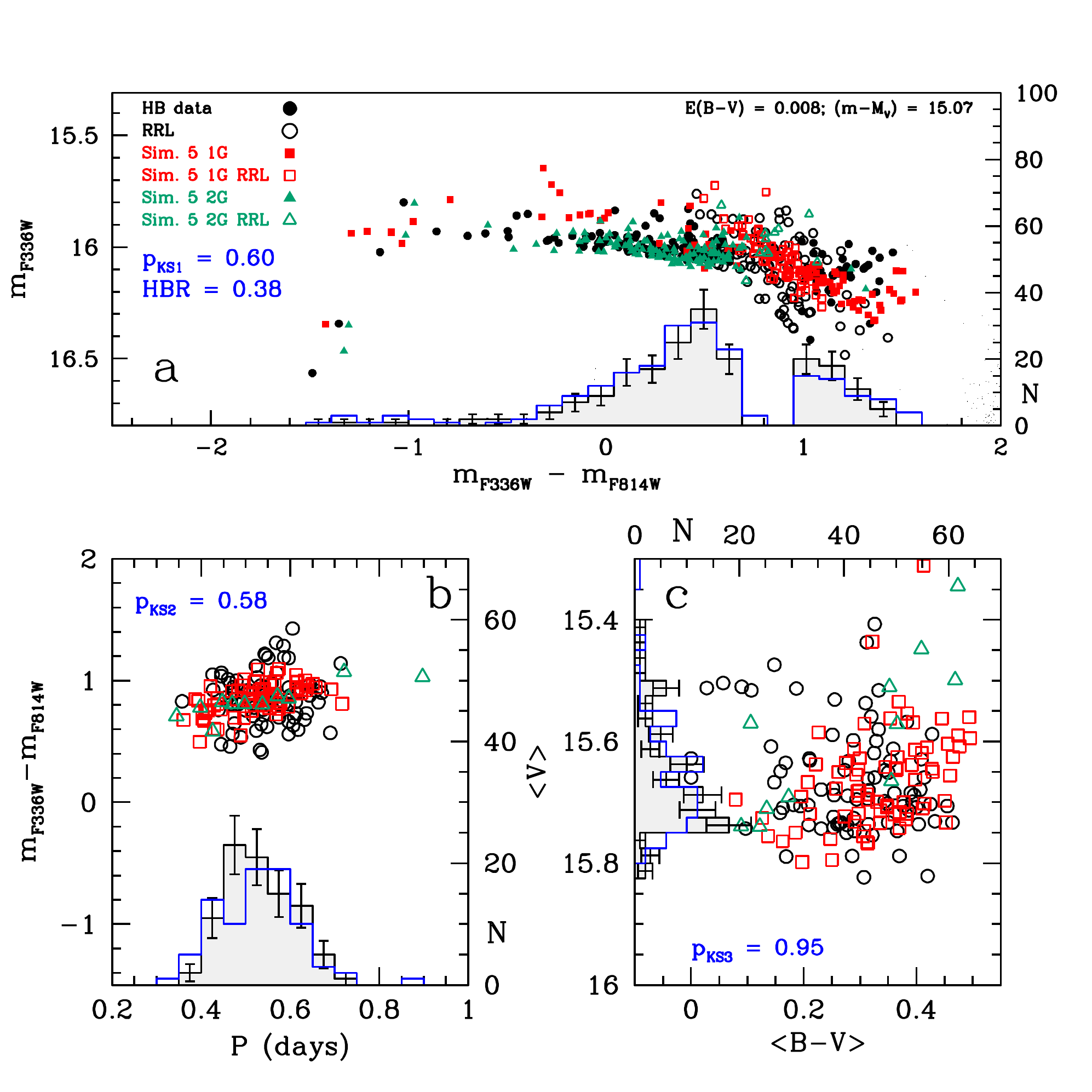}
\caption{
As Figure \ref{PIC_results3} but for Sim. 5. 
}
\label{PIC_results7}
\end{figure*}

\begin{figure*}
\hspace{-0.5cm}
\includegraphics[trim={1.8cm 0cm 0 0cm},width=2.2\columnwidth]{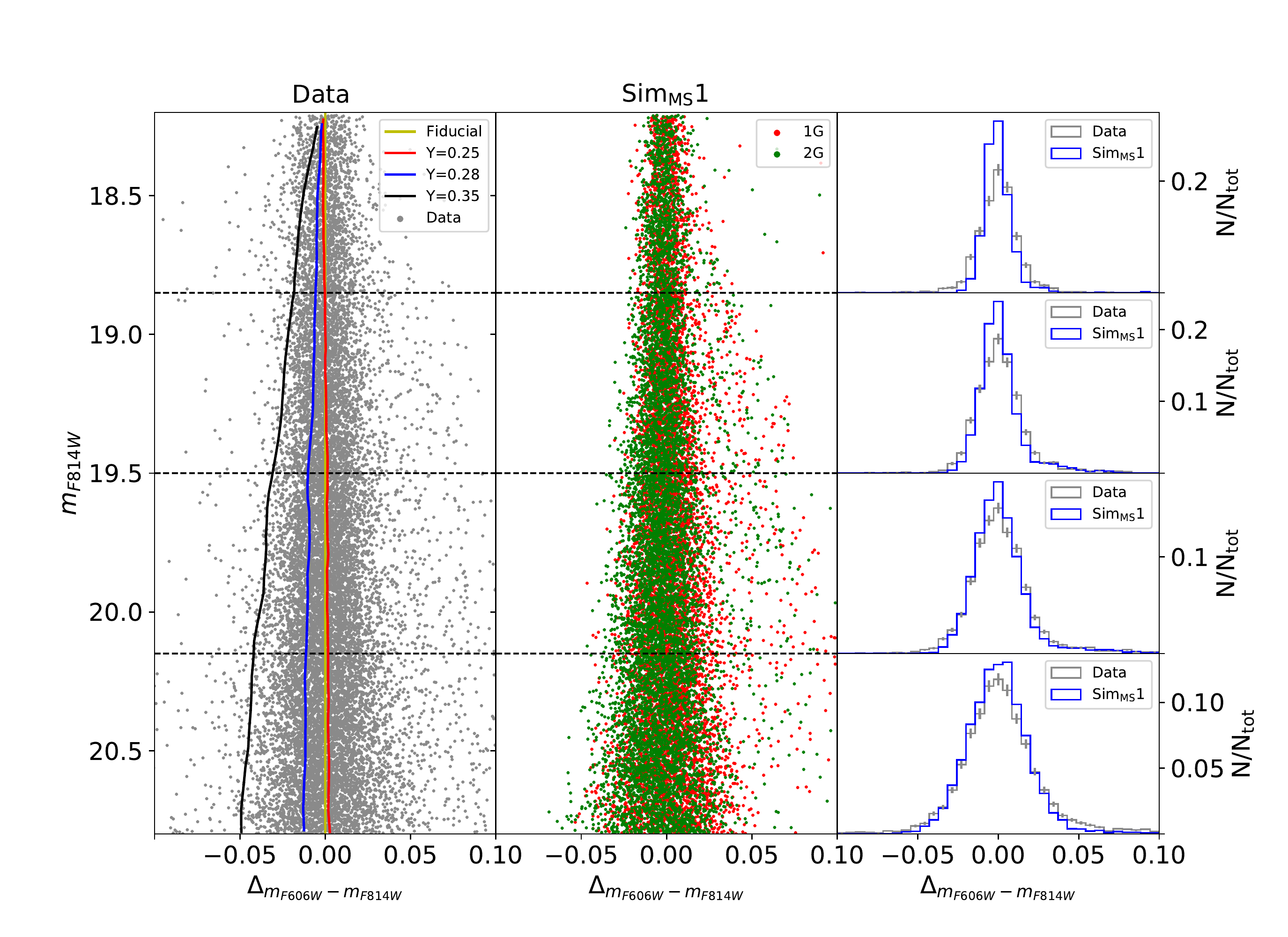}
\caption{
\textit{Left and central panels:} The comparison of the MS counterparts of Sim.1, dubbed $\rm Sim_{MS}1$, with the MS data of M\,3. The yellow line in the panel represents the fiducial of the data, while the red, blue and black ones are the isochrones with Y=0.25,0.28 and 0.35 respectively. \textit{Right panel:} The histograms compare the $\rm \Delta_{m_{F606W}-m_{F814W}}$ distribution of the data, grey,  and $\rm Sim_{MS}1$, blue in the four intervals of magnitudes highlighted in the panel.
}
\label{PIC_results8}
\end{figure*}

\begin{figure*}
\hspace{-0.5cm}
\includegraphics[trim={1.8cm 0cm 0 0cm},width=2.2\columnwidth]{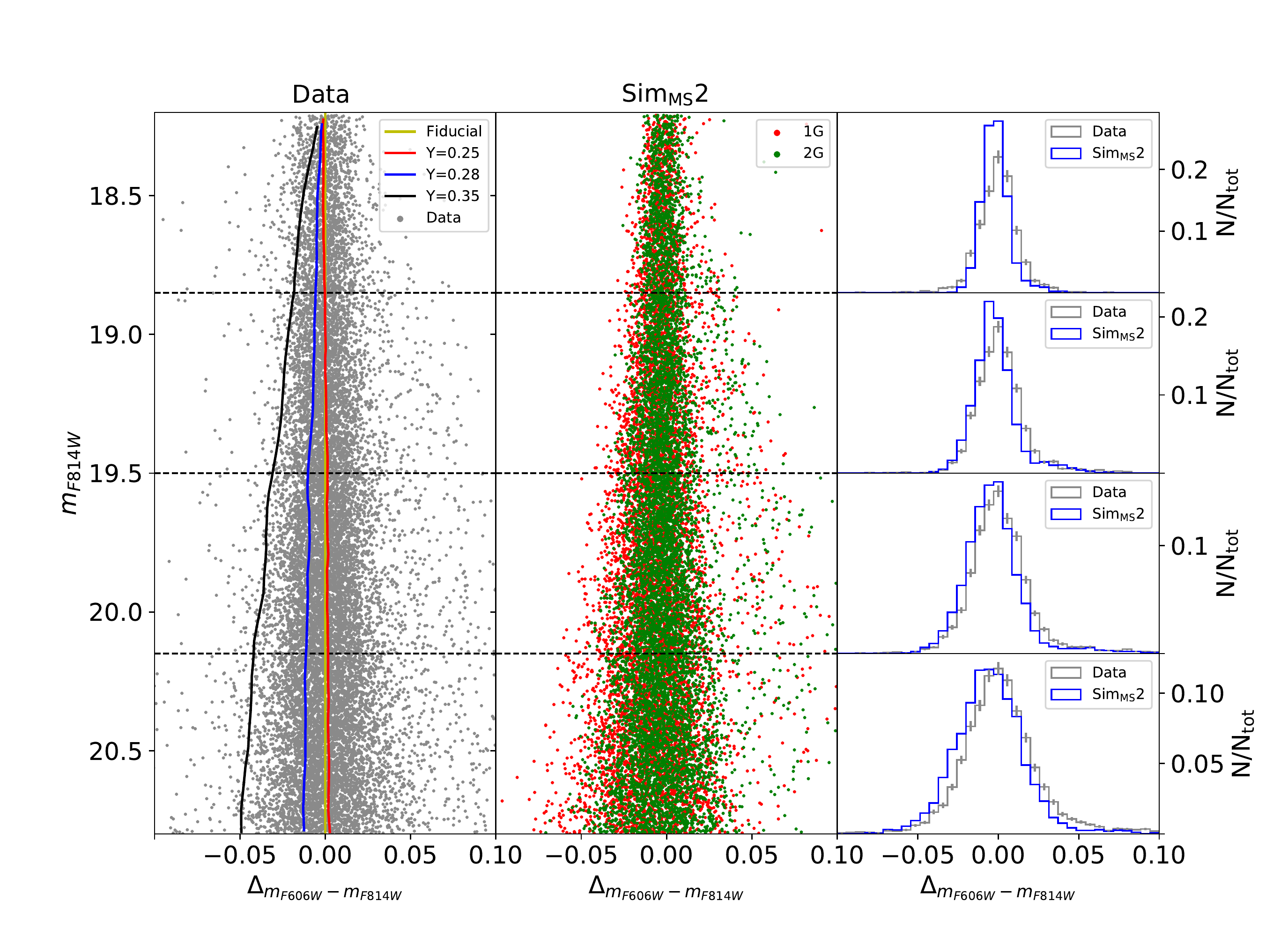}
\caption{
As Fig. \ref{PIC_results8} but for the MS counterpart of Sim. 2 ($\rm Sim_{MS}2$)
}
\label{PIC_results9}
\end{figure*}

\section{Results: simulations with no helium spread in the 1G}
\label{SEC_simres}

We start examining the case where the 1G has a very small or no helium spread. The observational constraints are summarized as follows:
\begin{itemize}

\item The mean and the max value of $\rm \Delta Y$ between the 1G and the 2G stars as obtained  by \citet[Table 4]{milone_2018}: $\rm \Delta Y_{2G,1G}=0.016\pm 0.005$ and $\rm \Delta Y_{2G,1G}^{max}=0.041\pm 0.009$ .

\item The colour distribution of HB stars in M3, as seen in Fig. \ref{PIC_results1}b.

\item The value of HBR as calculated from the \textit{HST} sample: $0.313\pm 0.034$.

\item The period distribution of our sample of RRL as extracted from \cite{benko_2006} and reported in Fig.\ref{PIC_results1}c.

\item The V magnitude distribution the RRL as extracted from \cite{benko_2006} (see Fig. \ref{PIC_results1}d). 

\end{itemize} 

We first examine several simulations containing only 1G stars to locate them on the HB. Given that both their helium mass fraction ($\rm Y\sim 0.25$) and its spread ($\rm \sigma^Y\sim 0.00$) can be constrained, the only remaining free parameters are the values of mass loss and its spread ($\rm \mu_{1G}; \sigma^{\mu}_{1G}$), which are constrained with the help of the simulation grid described in \S~\ref{SEC_models}.  We locate these stars on the red HB, and further constrain the values of these parameters with the RRL period and mean magnitude distributions. 

We now repeat the process including the 2G stars. The simple two populations model we adopted for the M4 case \citep{tailo_2019} does not work here as the morphology of the HB locus is more complex. We then need to split the blue HB stars in few groups ($\rm 2G_A,\, 2G_B,\, 2G_C$) with the 'A' group hosting the majority ($\sim 90\%$) of them. The helium abundances of these three groups are chosen in agreement to the $\rm \Delta Y_{2G,1G}$ and $\rm \Delta Y_{2G,1G}^{max}$ values mentioned before: we give the highest value to the bluest group of stars ($\rm 2G_C$) and we assign to the $\rm 2G_A$ group a value close to $\rm \Delta Y_{2G,1G}$ measured in \cite{milone_2018}. Finally we give to the $\rm 2G_B$ group an intermediate value. The final helium mass fraction values are listed in Table \ref{TAB_inputs}. 

Once the values of Y are set, the only free parameters remaining are the mass loss of the 2G stars and its spread  ($\rm \mu_{2G}; \sigma^{\mu}_{2G}$, for each group). We constrain these values by iteratively comparing the data with a new simulation grid which includes these new groups. 

Our best fit simulation (Sim. 1) is the one where $\rm \mu_{1G}=0.188\ M_\odot$. The mass loss of the 2GA, 2GB and 2GC groups is 0.016, 0.042 and $0.052\ M_\odot$ larger, respectively. The other inputs used are reported in (see Table \ref{TAB_inputs}). This confirms our previous findings in \cite{tailo_2019}.

 We apply the same procedure we used in \cite{tailo_2019} to evaluate the errors on these estimates and we obtain  $\rm \delta\mu_{1G,2GA} = 0.016 \pm 0.007 M_\odot$, $\rm \delta\mu_{1G,2GB}  = 0.032\pm0.009 M_\odot$ and $\rm \delta\mu_{1G,2GC}  = 0.052\pm0.013 M_\odot$ The results of Sim.\,1 are represented in Figure \ref{PIC_results3}; we also report some general parameters of this simulation in Table \ref{TAB_results}. 

Figure \ref{PIC_results3}a describes the simulated HB in the F336W and F814W CMD. The filled black dots indicate the observations while the open black dots indicate the variables. The simulation points are colour coded to distinguish the 1G and the 2G stars (red squares and the green triangles, respectively). The simulated RRL variables are represented by the open squares and triangles. The blue histogram is the colour distribution of the simulated, non variable, HB stars. The comparison with the homologous distribution for the data, the black shaded histogram, shows a good agreement, confirmed by the high probability value obtained from the KS  test ($\rm p_{KS1}$, see Table \ref{TAB_results}) for the two series of points. Furthermore, we obtain $\rm HBR = 0.322$, in agreement with the observed value. 
Note that the mass distribution of the 1G and 2G stars is different. The 1G stars have, on average, higher masses than the 2G ones (see Table \ref{TAB_results}). In addition the value of $\rm \delta \bar{Y}_{2G}$ is in agreement with the value given in \S~\ref{SEC_data}.

Sim.1 contains $\sim 100$ RRL variables, a number compatible with the observed sample. Fig. \ref{PIC_results3}b shows that the observed and simulated period distribution are in good agreement, also testified by the high probability value we obtain from the KS test ($\rm p_{KS2}$) and the similar mean period value (see Table \ref{TAB_results}). A good agreement is also found for the mean magnitude distribution of the variables (Fig. \ref{PIC_results3}c), again confirmed by the high value of the KS test p-value ($\rm p_{KS3}$).  

In summary, we have shown that a standard description of the HB in M\,3, assuming a single--valued helium content for 1G stars, and a helium enhancement consistent with the observational results describes in a satisfactory way the colour distribution and the RRL variables, as found in previous work on this cluster \citep[e.g.][]{catelan_2004,castellani_2005,caloi_2008,denissenkov_2017}, although adopting slightly different input parameters (see \S~\ref{SEC_Concl}). 

\section{Introducing the helium spread in 1G stars}
\label{SEC_sim2}

We now add a new constraint to the simulations: an internal helium spread among 1G stars. The spread has been computed by using the ChM map of 1G stars, by assuming that their distribution along the ($\rm m_{F275W}-m_{F814W}$)  colour  (see Fig. \ref{PIC_results1}a) is due to helium variation. 

We assign a standard helium value to those stars with $\rm \Delta_{F275W,F814W}> -0.03$. This choice is made following the indication from the error distribution in Figure 8 in \citet[][and references therein]{milone_2018}. We then assign the maximum value measured in \cite{milone_2018}, $\sim 0.1$, to the bluest star in the 1G sample, located at  $\rm \Delta_{F275W,F814W}\sim -0.4$. We linearly interpolate the helium value for the stars with $\rm -0.4 < \Delta_{F275W,F814W}< -0.03$. The distribution is shown in Figure \ref{PIC_results4}, where it is compared with the analogous distribution for the 2G stars; $\sim 30\%$ of the 1G stars preserve a standard helium content. 
We now attempt to achieve a new fit of the HB by including this 1G helium spread. The inputs of the simulations from 2 to 5 are reported in Table \ref{TAB_inputs} and some of their general properties in Table \ref{TAB_results}.

\subsection{Sim.\,2: including the 1G helium spread into Sim.\,1}
Figure \ref{PIC_results5} shows what happens by simply including the 1G helium spread, but leaving unaltered the other inputs of Sim.\,1. Now $\rm HBR\sim 0.649$, significantly higher than the observed value  because a good fraction of the of red HB stars goes to overpopulate the blue region; also the KS test confirms a very low probability value. In fact, the stars with larger helium have smaller masses and occupy bluer positions along the HB, as we have left the mass loss of all the 1G stars unaltered (see Table \ref{TAB_results}).

Sim. 2 has $\sim 70$ RRL variables, a number too small to be compatible with the observed sample. The period distribution is described in Fig. \ref{PIC_results5}b. The histograms in the panel testify that the simulated distribution is not a good description of the observed one for it is too wide. The discrepancy between the simulated and the observed variable stars is also indicated by the different distribution of their mean magnitudes (see Table \ref{TAB_results}). Fig. \ref{PIC_results5}c describes, as in the previous case, the distribution in $\rm M_V$ for the two samples. We see that the simulated ones do not have the peak we observe; the discrepancy is also confirmed by the low KS probability value  (see Table \ref{TAB_results}). 

To further explore how Sim. 1 changes, we realize a simulation using a distribution with a lower maximum, i.e. the one suggested by the findings in \cite{lardo_2018}. We obtain a simulated HB with $\sim$100 RRL and HBR$\sim$0.46; the latter is too high to be considered in agreement with the observed value. Furthermore, the low probability values we obtain from the KS tests for this new simulation (Sim. 2.2 in Table \ref{TAB_results}) indicate a strong discrepancy with the observations also in this case. 

\subsection{Sim.\,3: lowering the 1G mass loss}
Since one of the main issues of Sim. 2 is the low number of red HB stars, we test if a simulation with a lower mass loss value, for now arbitrarily chosen, would work.  Sim.\,3 adopts $\rm \mu_{1G}=0.175 M_\odot$ and the other inputs listed in Table \ref{TAB_inputs}. The rHB stars are better reproduced, but we are still overpopulating the blue HB, obtaining a $\rm HBR = 0.517$. The number of RRL variables is even smaller than in Sim.2, $\sim 45$, thus Sim.\,3 is not a good description of the observations, and in fact the KS tests report low probability values(see Table \ref{TAB_results}). 
If we lower the $\rm \mu_{1G}$ even more, the HBR value would be lowered as well, but the number of RRL variables would become lower too, thus representing a worse description. In the same manner an higher value of $\rm \mu_{1G}$ would produce even higher values of HBR, producing a worse simulation as well. We therefore have to reject this simulation, and try a different approach.

\subsection{Sim.\,4: starting the simulation from the variable stars distribution}
\label{SEC_sim3}
Because the majority ($\sim 80\%$) of the RRL variable occupies the lower luminosity part of the strip (Fig. \ref{PIC_results1}d), we can assign them ($\sim 75$) to the standard helium part of the 1G. We then give to the remaining stars in the 1G of Sim.1 ($\sim 55$) the helium spread previously derived (Figure \ref{PIC_results4}).  We leave the 2G inputs untouched: this gives us an additional $\sim 15$ RRL variables for a total of $\sim 90$ RRL, a number compatible with the observed sample. 

By using the same procedure adopted for Sim.1, we obtained the values of $\rm \mu_{1G}^{RRL}$ and $\rm \sigma_{1G}^{\mu^{RRL}}$ necessary to reproduce the features of the observed RRL variables. We then use these values to simulate the rest of the 1G stars. We obtain  a good fit of the variables with $\rm \mu_{1G}^{RRL} = 0.189 M_\odot$. This result is unsurprisingly similar to the value we obtain in Sim.1; due to the fact that the RRL features are among our main constraints.  The other inputs used in this simulation (Sim.\,4) are listed in Table \ref{TAB_inputs} and the results reported in Figure \ref{PIC_results6}, which follows the format of the previous ones.  

Figure \ref{PIC_results6}b and  \ref{PIC_results6}c confirm that the features of the RRL variable in Sim.4 are in a good agreement with the observations (by construction), as confirmed by  the high probability values of the KS tests reported Table \ref{TAB_results}.
Unfortunately, as shown in  Figure \ref{PIC_results6}a, with these values of $\rm \mu_{1G}^{RRL}$ and $\rm \sigma_{1G}^{\mu^{RRL}}$, the number of red HB stars is smaller than the observed one, while the blue HB is overpopulated ($\rm HBR=0.506$, still significantly higher than observed). The corresponding KS probability is $\rm p_{KS1}\sim 0.007$, see Table \ref{TAB_results}.

\subsection{Sim.\,5: lowering the mass loss of the red HB, with respect to Sim.\,4}
We attempt to improve the fit by lowering the mass loss of the non variable (NV) 1G stars. We repeat the procedure of Sim.1 including this new group of stars and find a good fit with $\rm \mu_{1G}^{NV} = 0.171\,M_\odot$. The other inputs of this new simulation (Sim.5) are reported in Table \ref{TAB_inputs}. Figure \ref{PIC_results7} reports its results.

As in the case of Sim. 4 we have that the RRL properties are well reproduced by construction (see Fig. \ref{PIC_results7}b, Fig. \ref{PIC_results7}c and Table \ref{TAB_results}). In Fig. \ref{PIC_results7}a we report the simulated HB stars of Sim.5. Apparently, we have a good reproduction of the observed sample: we indeed obtain a high probability value from the KS test (see Table \ref{TAB_results}). We have $\rm HBR = 0.384$, a value slightly higher than the observed one. Additional considerations, however, suggest that this simulation is not acceptable when framed in the context of GC evolution and the spectroscopic observations of M3. Indeed, for the non variable 1G stars, we reduced $\rm \mu$ of a non negligible quantity ($\rm \Delta \mu \sim -0.018\ M_\odot$). The current scenarios for the 1G formation describe those stars formed in the same environment. It is then unlikely that stars in the same generation suffer such diverse mass loss.  

The need to reduce the mass loss can be avoided if we change the value of $\rm M_{Tip}$ between the RRL and the non variable stars. This can be achieved in two ways: assuming they are \textit{younger} or \textit{increasing their metallicity}. From our isochrones database we found that an age difference of $\rm \sim -1.0\ Gyr$ would produce the same effects; similarly with $\rm \delta [Fe/H] \sim 0.18$. Both these possibilities have to be excluded because, at the time of this writing, they are not supported by the current spectroscopic and photometric observational framework for M3.

Furthermore, there is no evidence that a significant variation of [C+N+O/Fe] has been directly observed within the 1G, see \cite{marino_2019} and \citet[][for the case of NGC\,2808]{cabrera_2019}. Thus as aforementioned, we then need to reject this simulation as unsatisfactory.

\section{The MS counterparts of the HB simulations}
\label{SEC_SimMS}

Now that we have a description of the stellar populations hosted in the HB of M\,3, we check if our findings are adequate to describe the MS as well. We analyse the MS counterpart of Sim.1 and Sim.2 and compare them to the MS data in the F606W and F814W bands. We will perform the simulations as previously described (see \S\,\ref{SEC_models}), dividing the total number of MS stars in the groups listed in Table \ref{TAB_inputs}. 

Figures \ref{PIC_results8} and \ref{PIC_results9} report the results of these new tests. The left panel of both figures reports the MS data of M\,3; the central panel describes these new simulations, $\rm Sim_{MS}1$ and $\rm Sim_{MS}2$, respectively for Fig. \ref{PIC_results8} and  \ref{PIC_results9}. In the panels, the yellow line represents the fiducial of the data. The red, blue and black lines are the isochrones with Y=0.25, 0.28 and 0.35, respectively. As in previous figures, the 1G  and the 2G are represented as red and green dots, respectively. The histograms on the right panels compare the $\rm \Delta_{m_{F606W}-m_{F814W}}$ distribution of both the data, grey, and the simulations, blue. 

We immediately see that in $\rm Sim_{MS}2$ the 1G overlaps with the 2G entirely, populating even the blue side of the MS. This is the direct consequence of its helium distribution. Indeed, in $\rm Sim_{MS}2$, the 1G stars have, on average, higher helium mass fraction values than the ones in $\rm Sim_{MS}1$ and their counterparts in the 2G (see Tab. \ref{TAB_results}).  The net result is that the blue side of the simulated MS is overpopulated compared to the observed one, as  the comparison of the histograms in the figures shows. This simple test reinforces the point we made with the HB simulations in the previous sections.

\section{Discussion and conclusions}
\label{SEC_Concl}

We present a collection of HB and MS simulations to investigate whether the tentative interpretation of the 1G colour extension observed in the ChM of M\,3 in terms of a pure helium spread \citep{milone_2015,lardo_2018,milone_2018}
is consistent with other observations of this cluster. 
The parameters of our simulations must be chosen in order to satisfy the following observational constraints\footnote{described in \S\,\ref{SEC_data}}:
\begin{itemize}
\item the mean and the max value of $\rm \Delta Y$ between the 1G and the 2G stars as obtained  by \citet[Table 4]{milone_2018};
\item the colour distribution of the HB stars in M3 and the value of the HB ratio of our sample;
\item the period and mean magnitude distributions of our RRL sample.
\end{itemize}
The parameters which can not be directly constrained from these observations, namely the mass loss of the 1G and the 2G star groups with their dispersion ($\rm \mu_{1G(2G)}$ and $\rm \sigma^\mu_{1G(2G)}$), have been constrained by iteratively comparing a series of  simulation grids with the data. We have thus shown only the most representative results of a large and deep exploration of the parameters space.

The main result of this work is that there is no combination of parameters that reproduces correctly all the observational constraints at the same time, \textit{if we include within the 1G stars the internal helium spread suggested by the most direct interpretation of the Chromosome Map}. We conclude that the reasons for the colour spread of the 1G remain to be understood.
We, instead, find that any helium spread associated with the 1G stars in Sim. 1 has to be very small (<0.002).
This also rules out the possibility that a smaller but non negligible helium enhancement \citep[as the one described by][]{lardo_2018} can be used to describe the 1G stars in this cluster.

The simulations we show in this work have other interesting features:

\begin{enumerate}
\item When the 1G stars have small or no helium spread, all the red HB stars and the majority of RRL variables belong to 1G. This result also confirms the approach already followed by different authors in the past \citep[e.g.][]{caloi_2008,denissenkov_2017}. The mass loss needed to reproduce the position on the HB of the 1G star locates them near the red border of the instability strip. 

\item We find that the mass loss spread for the two population has to be small ($\rm \sigma^\mu \sim 0.005 M_\odot$). This value is larger than the one from \cite{caloi_2008} ($\rm \sigma^\mu \lesssim 0.003\ M_\odot$), thanks to the lower peak of the period distribution of RRL variables. On the other hand, \cite{vandenberg_2016} and \cite{denissenkov_2017} found a larger value ($\rm \sigma^\mu \sim 0.008 M_\odot$ in their 1G). This is conseguence of adopting the catalogue from  \cite{cacciari_2005}, who are interested in the accurate photometry of these stars and thus exclude the variables which show the Blazhko type variability \citep{blaz_1907}  from the number versus period distribution of RRL. Although the colours of these stars are more difficult to be measured with accuracy, the periods of these RRL variables are well known and contained in the complete \cite{benko_2006} catalogue.  Discarding the Blazhko RRL ($\sim 30$\% of the sample), the number versus period distribution becomes flatter, and is consistent with a larger value of  $\sigma_\mu$.

\item Having both $\rm \Delta Y_{2G,1G}$ and $\rm \Delta Y_{2G,1G}^{max}$ fixed by the observations reported in \cite{milone_2018} we can also constrain the mass loss difference between the two populations. In Sim.\,1 we find$\rm \delta\mu_{1G,2GA} = 0.016 \pm 0.007 M_\odot$ and $\rm \delta\mu_{1G,2GC}  = 0.052 \pm 0.013 M_\odot$. The errors have been estimated with the procedure used in \cite{tailo_2019}. All the other simulations show similar differences. We found a similar trend in the HB of the cluster M\,4 \citep{tailo_2019}, hinting that this can be a common behaviour in GCs. \cite{denissenkov_2017} also found larger mass loss for the 2G stars in M\,3. This difference in the mass loss is also consistent with the very small ---if any--- helium increase across the RRL variables region, found by \cite{catelan_2009}, but does not rule out helium as a primary parameter to determine the morphology of the whole HB. 

\item We regard the difference in mass loss between 1G and 2G as a result of the different formation conditions of the two populations \citep[e.g. different initial rotation rates, as we suggested in][]{tailo_2015,tailo_2016b,tailo_2017,tailo_2019} . Consequently, we attribute the same mass loss to all 1G stars, even for the simulations with 1G helium spread. However, increasing values of mass loss for the helium rich stars in the 1G lead to even worse agreement with the observations, as the blue HB would include even more stars (thus obtaining an even higher HBR). 

\item The RRL variables in our simulations are made up almost entirely by 1G stars. In Sim.1 only 14 RRL ($\sim 15\%$) belong to the 2G. The 2G RRL are evolved HB stars, crossing the instability strip from the blue side and they are, on average, more luminous than their 1G counterparts.

\item Sim. 5 shows an almost adequate reproduction of the observed HB stars in M3, except for the value of HBR, which is slightly higher (0.384, see Fig. \ref{PIC_results7} and Table \ref{TAB_results}). To obtain this result, we needed to lower the mass loss of the non variable stars of a non negligible quantity ($\rm \Delta \mu \sim -0.018\ M_\odot$). At the time of this writing we can not justify this assumption within the current spectroscopic and photometric observational framework of M\,3 therefore we have to reject this last simulation. 

\item The simulations of the MS in Fig. \ref{PIC_results8} and \ref{PIC_results9}, counterparts of Sim.1 and Sim.2,  show that the blue MS would be overpopulated if the 1G includes a large helium spread, reinforcing the results obtained from the HB simulations. 

\end{enumerate}

In conclusion, the exam of the CMD data of M\,3, in particular its HB stellar distribution, the period distribution of its RR\,Lyrae variable stars, and the color distribution of the main sequence stars,  do not support the hypothesis that the color spread of the 1G is due to internal helium variations among this population, thus the problem remains to be understood.

Notice that M\,3 and M\,13 are the most typical example of "second parameter" pair \citep[e.g. see the now classic work from][]{caloi_2005}, and, while M\,3 has an extended 1G, M\,13 has a very compact 1G \citep[see the reproduction of their ChM in][]{milone_2017,milone_2018}. Is the 1G extension  part of the (still unsolved) "second parameter" problem?

\section*{Acknowledgements}
This work has received support from the European Research Council (ERC) via the European Union's Horizon 2020 research innovation programme (Grant Agreement ERC-StG 2016, No 716082 'GALFOR', PI: Milone), and the European Union's Horizon 2020 research and innovation programme via the Marie Sklodowska-Curie (Grant Agreement No 797100, beneficiary: Marino). M.T. and A.P.M. acknowledge support from MIUR through the the FARE project R164RM93XW ‘SEMPLICE’ (PI: Milone).

\bibliographystyle{mnras}
\bibliography{m3} 




\bsp    
\label{lastpage}
\end{document}